\documentclass[noshowpacs,amsmath,
twocolumn,
superscriptaddress,
8pt,
aps,prb]{revtex4-1}  
\usepackage{setspace}
\usepackage{amsmath}
\usepackage{graphicx,afterpage}
\usepackage[nearskip,margin = 0pt]{subfig}
\usepackage{verbatim}  
\usepackage{amsfonts}
\usepackage{amssymb}

\begin{document}

\title{Ultra-high-Q wedge-resonator on a silicon chip}

\author{Hansuek Lee, Tong Chen, Jiang Li, Ki Youl Yang, Seokmin Jeon, Oskar Painter, and Kerry J. Vahala\\
T. J. Watson Laboratory of Applied Physics, California Institute of Technology, Pasadena, California 91125, USA}

\date{\today}

\maketitle

\noindent
\textbf{Ultra-high-Q optical resonators are being studied across a wide range of research subjects including quantum information, nonlinear optics, cavity optomechanics, and telecommunications \cite{vahala_nature,review_vahala,review_vahala2,review_ilchenko,review_ilchenko2,review_comb,cqed}. Here, we demonstrate a new, resonator on-a-chip with a record Q factor of 875 million, surpassing even microtoroids \cite{toroid}. Significantly, these devices avoid a highly specialized processing step that has made it difficult to integrate microtoroids with other photonic devices and to also precisely control their size.  Thus, these devices not only set a new benchmark for Q factor on a chip, but also provide, for the first time, full compatibility of this important device class with conventional semiconductor processing. This feature will greatly expand the possible kinds of system on a chip functions enabled by ultra-high-Q devices.}\\
\indent
\begin{figure*}[!bt]
  \begin{centering}
  \includegraphics[trim = 65mm 65mm 45mm 60mm, clip, width=17.5cm]{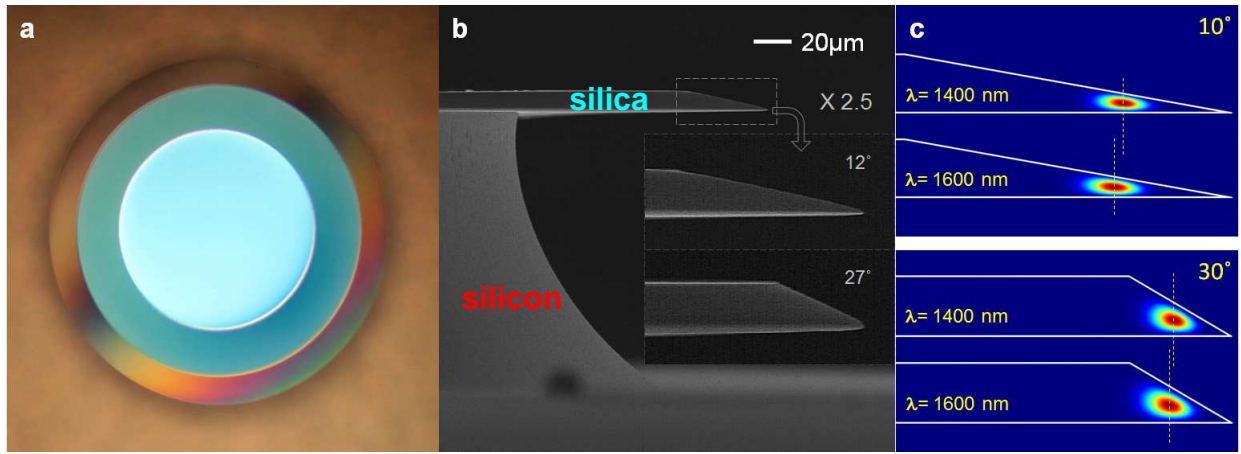}
  \captionsetup{singlelinecheck=no, justification = RaggedRight}
  \caption{\textbf{Micrographs and mode renderings of the wedge resonator from top and side views.} \textbf{(a)} An optical micrograph shows a top view of a $1\,$mm diameter wedge resonator. \textbf{(b)} A scanning electron micrograph shows the side-view of a resonator. The insets here give slightly magnified micrographs of resonators in which the wedge angle is $12$ degrees (upper inset) and $27\,$ degrees (lower inset). \textbf{(c)} A rendering shows calculated fundamental mode intensity profiles in 10 degree and 30 degree wedge angle resonators at two wavelengths. As a guide, the center-of-motion of the mode is provided to illustrate how the wedge profile introduces normal dispersion that is larger for smaller wedge angles. }
  \label{fig1}
\end{centering}
\end{figure*}
Long photon storage time (high Q factor) in microcavities relies critically upon use of low absorption dielectrics and creation of very smooth (low scattering) dielectric interfaces. For chip-compatible devices, silica has by far the lowest intrinsic material loss. Microtoroid resonators combine this low material loss with a reflow technique in which surface tension is used to smooth lithographic and etch-related blemishes \cite{toroid}. At the same time, reflow smoothing makes it very challenging to fabricate larger diameter UHQ resonators and likewise to leverage the full range of integration tools and devices available on silicon. The devices reported here attain ultra-high-Q performance using only conventional semiconductor processing methods on a silicon wafer. Moreover, the best Q performance occurs for diameters greater than $500$ microns, a size range that is difficult to access for microtoroids on account of the limitations of the reflow process. Microcombs will benefit from such a combination of UHQ and larger diameter resonators (microwave-rate free-spectral-range) to create combs that are both efficient in turn-on power and that can be self-referenced \cite{review_comb}. Moreover, integrated reference cavities and ring gyroscopes are two other applications that can benefit from larger ($1-50\,$mm diameter) UHQ resonators. Fabrication control of the free-spectral range to $1:20,000$ is also demonstrated here, opening the possibility of precision repetition rate control in microcombs or precision spectral placement of modes in certain nonlinear oscillators \cite{SBS_carmon,SBS_ivan}. \\
\indent
Earlier work considered the Q factor in a wedge-shaped resonator fabricated of silica on a silicon wafer. Q factors as high as $50$ million were obtained \cite{PRA}. That approach isolated the mode from the lithographic blemishes near the outer rim of the resonator by using a shallow wedge angle. In the current work, we have boosted the optical Q by about $20$X beyond these earlier results through a combination of process improvements. These improvements make it unnecessary to isolate the mode from the resonator rim. Indeed, the highest Q factor demonstrated uses the largest wedge angles. A top-view optical micrograph is provided in figure $1$ to illustrate the basic geometry. The process flow begins with thermal oxide on silicon, followed by lithography and oxide etching with buffered hydrofluoric acid. In the insets to figure $1$, scanning electron micrographs of devices featuring  $12$-degree and $27$-degree wedge angles are imaged. Empirically, the angle can be controlled through adjustment of the photoresist adhesion using commercially available adhesion promoters. The oxide disk structures function as an etch mask for an isotropic dry etch of the silicon using XeF$_2$. During the dry etch, the silicon undercut is set so as to reduce coupling of the optical mode to the silicon support pillar. This value is typically set to about $100$ microns for $1\,$mm diameter structures and over $150$ microns for $7.5\,$mm diameter disks, however, smaller undercuts are possible while preserving ultra-high-Q performance.  Further information on the processing is given in the Methods section.\\
\indent
To measure intrinsic Q factor, devices were coupled to SMF-$28$ optical fiber using a fiber taper \cite{taper, ideality_PRL} and spectral lineshape data were obtained by tuning an external cavity semiconductor laser across the resonance while monitoring transmission on an oscilloscope. To accurately calibrate the laser scan in this measurement, a portion of the laser output was also monitored after transmission through a calibrated Mach-Zehnder interferometer having a free spectral range of $7.75$ MHz. The inset in figure $2$ shows a spectral scan obtained on a device having a record Q factor of $875$ million. In these measurements, the taper coupling was applied on the upper surface of the resonator near the center of the wedge region. Modeling shows that the fundamental mode has its largest field amplitude in this region. Moreover, this mode is expected to feature the lowest overall scattering loss resulting from the three, dielectric-air interfaces as well as from the silicon support pillar. An additional test that can be performed to verify the fundamental mode is to measure the mode index by monitoring the free-spectral-range (FSR). The fundamental mode features the largest mode index and hence smallest FSR.\\
\indent
\begin{figure}[tb!]
\hspace{-.5cm}
  \includegraphics[width=8.5cm]{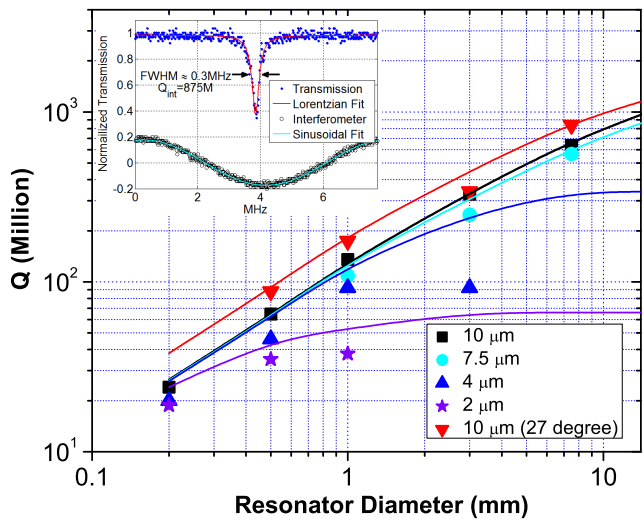}
  \captionsetup{singlelinecheck=no, justification = RaggedRight}
\caption{\textbf{Data showing the  measured Q factor plotted versus resonator diameter with oxide thickness as a parameter.} The solid lines show the predicted Q factor from a model that accounts for surface roughness induced scattering loss and also material loss. The rms roughness is measured using an AFM (see Methods section for values) and the fitted bulk material loss corresponds to a Q value of $2.5$ billion. The red data points correspond to a wedge angle of $27$ degrees. All other data are obtained using a wedge angle of approximately $10$ degrees. The inset shows a spectral scan for the case of a record Q factor of $875$ million. The sinusoidal curve accompanying the spectrum is a calibration scan performed using a fiber interferometer.}
\label{fig:spiral}
\end{figure}
The typical coupled power in all measurements was maintained around $1$ microWatt to minimize thermal effects.  However, there was little or no evidence of thermal effects in the optical spectrum. Typically, these appear as an asymmetry in the lineshape and also a scan-direction dependent (to higher or lower frequency) spectral linewidth. As a further check that thermal effects were negligible, ring down measurements \cite{ring_down} were also performed on a range of devices for comparison to the spectral-based Q measurement. For these, the laser was tuned into resonance with the cavity and a lithium niobate modulator was used to abruptly switch off the input. The output cavity decay rate was then monitored to ascertain the cavity lifetime. Ring-down data and spectral linewidths were consistently in good agreement. This insensitivity to thermal effects is a result of the larger mode volumes of these devices in comparison to earlier work on microtoroids (for which thermal effects must be carefully monitored).  The mode volumes in the present devices are typically $100-1000$X larger.\\
\indent
\begin{figure}[b!]
\hspace{-.5cm}
  \begin{centering}
  \includegraphics[width=8.5cm]{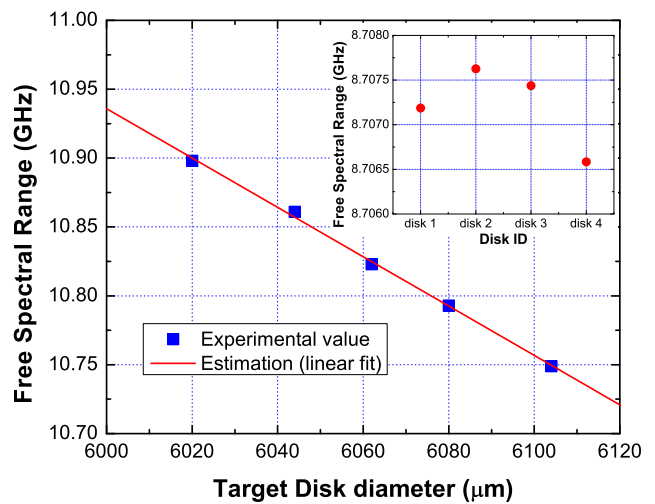}
\end{centering}
\captionsetup{singlelinecheck=no, justification = RaggedRight}
  \caption{\textbf{Plot of measured free spectral range (FSR) versus the target design-value resonator diameter on a lithographic mask.} The plot shows one device at each size and five different sizes. The rms variance is $2.4\,$MHz (relative variance of less than $1:4,500$). The inset shows the FSR data measured on four devices having the same target FSR. An improved variance of $0.45\,$MHz is obtained (a relative variance of $1:20,000$).}
  \label{fig3}
\end{figure}
Measurements showing the effects of oxide thickness and device diameter on Q factor are presented in the figure $2$ main panel. Four, oxide thicknesses are shown ($2$, $4$, $7.5$ and $10$ microns) over diameters ranging from $0.2\,$mm to $7.5\,$mm. All data points, with the exception of the red points, correspond to a wedge angle of approximately $10$ degrees. The upper most (highest Q at a given diameter) data correspond to a wedge angle of $27$ degrees. The solid curves are a model of optical loss caused by surface scattering on the upper, wedge, and lower oxide-air interfaces and by bulk-oxide loss. In the model, the surface roughness was measured independently on each of these surfaces using an atomic force microscope (AFM) (r.m.s. roughness values are given in the Methods section). The bulk optical loss of the thermal silica corresponds to a Q value of $2.5$ billion by fit to the data. The data corresponding to the $10$ degree wedge angle show that Q increases for thicker oxides and also larger diameters. Using the model, this trend can be understood to result from loss that is caused primarily by scattering at the oxide-air interfaces. Specifically, both thicker oxides and larger diameter structures feature a reduced field amplitude at the dielectric-air interface, leading to reduced scattered power. A slight, overall boost to the Q factor is possible by increasing the wedge angle. In this case, the mode experiences reduced upper and lower surface scattering as compared to the smaller angle case. A record Q factor of $875$ million for any chip based resonators is obtained under these conditions. In general, there is reasonably good agreement between the model and the data, except in the case of the thinner oxides. For these thinner structures, there is a tendency for stress-induced buckling to occur at larger radii. This is believed to create the discrepancy with the model. \\
\indent
\begin{figure}[tb!]
  \begin{centering}
  \includegraphics[width=8.5cm]{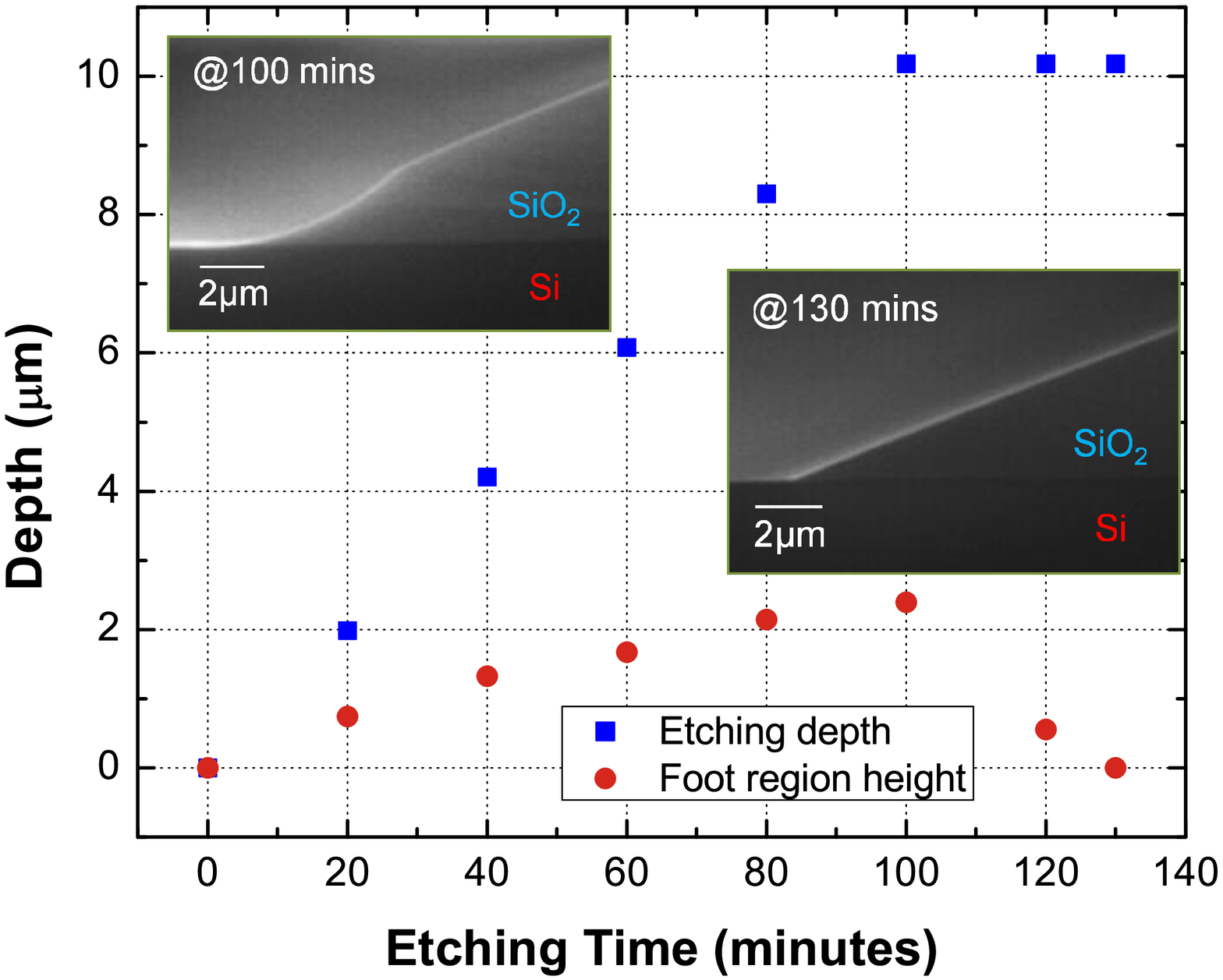}
\end{centering}
\captionsetup{singlelinecheck=no, justification = RaggedRight}
  \caption{\textbf{Data plot showing the effect of etch time on appearance of the ``foot'' region in etching of a $10$ micron thick silica layer.} The foot region is a separate etch front produced by wet etch of silica that is empirically observed to adversely affect the optical Q factor. The data show that by control of the etch time the ``foot'' region can be eliminated. The upper-left inset is an image of the foot region and the lower right inset shows the foot region eliminated by increase of the wet etch time. }
  \label{fig4}
\end{figure}
The ability to lithographically define ultra-high-Q resonators as opposed to relying upon the reflow process enables a multi-order-of-magnitude improvement in precision control of resonator diameter and FSR. This feature is especially important in microcombs and also certain nonlinear sources \cite{SBS_carmon, SBS_ivan}. As a preliminary test of the practical limits of FSR control, two studies were conducted. In the first, a series of resonator diameters were set in a CAD file used to create a photo mask. A plot of the measured FSR (fundamental mode) versus CAD file target diameter is provided in figure $3$ (main panel). The variance from ideal linear behavior is $2.4\,$MHz, giving a relative variance of better than $1:4,500$ (FSR $\approx$ $11$ GHz). The inset to figure $3$ shows that for separate devices having the same target CAD file diameter, the variance is further improved to a value of $0.45\,$MHz or $1:20,000$.\\
\indent
The Q factor for these new resonators is not only higher in an absolute sense than what has been possible with microtoroids, but it also accesses an important regime of resonator FSR that has not been possible using microtoroids. To date, the smallest FSR achieved with the toroid reflow process has been $86\,$GHz (D $=750\,\mu m$) and the corresponding Q factor was $20$ million \cite{kippenberg_PRL_2008}. The present structures attain their best Q factors for FSRs that are complementary to microtoroids (FSRs less than $100\,$GHz). This range has become increasingly important in applications like microcombs where self-referencing is important. Specifically, low turn-on power and microwave-rate repetition are conflicting requirements in these devices on account of the inverse dependence of threshold power on FSR. However, such increases can be compensated using ultra-high-Q because turn-on power depends inverse quadratically on Q \cite{OPO_vahala}. The ability to manipulate normal dispersion through the wedge angle (see figure 1) can be shown to provide control over the zero dispersion point in spectral regions where silica exhibits anomalous dispersion. Ultra-high-Q performance in large area resonators is also important in rotation sensing \cite{rotation} and for on-chip frequency references \cite{freq_ref, freq_ref2}. In the former case, the larger resonator area enhances the Sagnac effect. In the latter, the larger mode volume lowers the impact of thermal fluctuations on the frequency noise of the resonator \cite{noise}. The precision control of FSR is important to determine repetition rate in microcombs, and also in applications such as stimulated Brillouin lasers where a precise match of FSR to the Brillouin shift is a prerequisite for oscillation. Application of these devices to low turn-on power, microwave-rate microcombs and to high-efficiency SBS lasers will be reported elsewhere. Finally, an upper bound to the material loss of thermal silica was established in this work. The value of $2.5$ billion bodes well for further application of thermal silica to photonic devices. \\
\newline
\noindent
\textbf{Methods}\\
Disks were fabricated on ($100$) prime grade float zone silicon wafers. Photo-resist was patterned using a GCA $6300$ stepper on thermally grown oxide of thickness in the range of $2-10$ microns. Post exposure bake followed in order to cure the surface roughness of photo-resist pattern which acted as an etch mask during immersion in buffered hydrofluoric solution (Transene, buffer-HF improved). Careful examination of the wet etch revealed that the vertex formed by the lower oxide and upper surface contains an etch front that is distinct from that associated with the upper surface (see ``foot'' region in figure $4$ inset). This region has a roughness level that is higher than any other surface and is a principle contributor to Q degradation. By extending the etch time beyond what is necessary to reach the silicon substrate, this foot region can be eliminated as shown in figure $4$. With elimination of the foot etch front, the isotropic and uniform etching characteristic of buffered hydrofluoric solution results in oxide disks and waveguides having very smooth wedge-profiles which enhance Q factors. After the conventional cleaning process to remove photo-resist and organics, silicon was isotropically etched by xenon difluoride to create an air-cladding whispering gallery resonator. An atomic force microscope was used to measure the surface roughness of the three, silica-air dielectric surfaces. For the lower surface, the resonators were detached by first etching the silicon pillar to a few microns in diameter and then removing the resonator using tape. The r.m.s. roughness values on $10$-degree wedge-angle devices are: $0.15\,$nm (upper), $0.46\,$nm (wedge), $0.70\,$nm (lower); and for $27$-degree wedge-angle devices are: $0.15\,$ nm (upper), $0.75\,$nm (wedge), $0.70\,$nm (lower). The correlation length is approximately a few hundred nm. The difference in the wedge surface roughness obtained for the large and small wedge angle cases is not presently understood. \\

\hspace{10 mm}

\textbf{Acknowledgments}
We gratefully acknowledge the Defense Advanced Research Projects Agency under the iPhod and Orchid programs and also the Kavli Nanoscience Institute at Caltech. H. L. thanks the Center for the Physics of Information.
\hspace{10 mm}

%
\hspace{10 mm}

\bibliography{ref}
\end{document}